\documentstyle[twocolumn,aps,epsfig]{revtex}

\draft

\def\lapproxeq{\lower .7ex\hbox{$\;\stackrel{\textstyle <}{\sim}\;$}}

\def\Ecut{E_c}

\begin{document}

\title{Neutrinos from propagation of ultra--high energy protons}

\author {Ralph~Engel, David~Seckel and Todor~Stanev}
\address {
Bartol Research Institute, 
University of Delaware, Newark, DE 19716, USA
}

\widetext
\wideabs{
\maketitle
\vspace*{5truemm}
We present a calculation of the production of neutrinos during 
propagation of ultra-high energy cosmic rays as may be produced in 
astrophysical sources. Photoproduction interactions are modeled with the 
event generator SOPHIA that represents very well the experimentally 
measured particle production cross sections at accelerator energies.  We 
give the fluxes expected from different assumptions on cosmic ray source 
distributions, cosmic ray injection spectra, cosmological evolution of 
the sources and different cosmologies, and compare them to the 
Waxman-Bahcall limit on source neutrinos. We estimate rates for detection 
of neutrino induced showers in a km$^3$ water detector. The ratio of the 
local high energy neutrino flux to the ultra-high energy cosmic ray flux 
is a crucial parameter in distinguishing between astrophysical and 
cosmological (top-down) scenarios of the ultra-high energy cosmic ray 
origin.} 

\date{Submitted to Phys Rev D, ???? 2000}
%%}

\narrowtext

\section{Introduction}
%%%%%%%%%%%%%%%%%%%%%%%%%%%%%%%%%%%%%%%%%%%%%%%%%%%%%%%%%%%%%%%%%%%%%%%%%%
  
 The highest energy cosmic rays are energetic enough to have
 photoproduction interactions on the microwave background. These
 collisions cause energy loss affecting the
cosmic ray spectrum~\cite{GZK} -- the GZK cutoff.
 In most astrophysical environments all secondary mesons produced
in photoproduction interactions decay into $\gamma$--rays
 and neutrinos. Shortly after the original papers on the GZK
 cutoff it was suggested~\cite{BerZat69} that guaranteed fluxes
 of ultra-high energy (UHE) neutrinos will be produced by the
 propagation of UHE cosmic ray (UHECR) protons in the Universe.
 This suggestion was followed by more sophisticated
 estimates~\cite{WTW,Stecker73,BerSmi75,BerZat77,Stecker79} that 
 attempted to predict more realistically the expected neutrino
 fluxes and relate the
 detection of such fluxes to the neutrino cross section at very high
 energy and the then unknown mass of the $W$ boson.
 Hill \& Schramm~\cite{HS85,HSW86} introduced the cosmological
 evolution of cosmic ray sources and used the measurements of
 the cosmic ray spectrum by the Haverah Park~\cite{Cunetal} and
 the Fly's Eye~\cite{FY85} experiments to determine minimum and
 maximum allowed 
 normalizations for the flux of such `propagation' neutrinos and
calculated detection rates for different types of detectors.
 More recent estimates include the work of Stecker and
 collaborators~\cite{SDSS}, Yoshida \& Teshima~\cite{YT93}, and 
 Protheroe \& Johnson~\cite{PJ00}. 
 
 Meanwhile the world statistics of ultra-high energy cosmic
 rays has significantly increased~\cite{NagWat00} and, most
 importantly, two events of energy substantially above 10$^{20}$~eV
 were detected by the Fly's Eye~\cite{FYHigh} and AGASA~\cite{AGASAhigh}
 experiments. These events suggest that the maximum energy
 of cosmic ray acceleration E$_{\rm max}$ may be significantly higher
 than the previous nominal estimate of 10$^{20}$ eV, if these
 events are not a result of the decay of extremely massive
 exotic particles~\cite{BKV} or other exotic processes~\cite{BatSigl00}.
 
 We assume that UHECR are of astrophysical origin and
 present here a new estimate of the expected neutrino fluxes
 generated during propagation by the ultra-high energy cosmic rays.
 We use recent results on the propagation of UHE protons~\cite{SEMPR}
 to estimate the neutrino production. The aim is to explore the
 neutrino  production with a photoproduction interaction model
 (SOPHIA~\cite{SOPHIA}) that fits well the experimentally measured
 multiparticle production data over a wide energy range. 
% The proton propagation calculation in
% Ref.~\cite{SEMPR} accounts for the scattering of cosmic rays in
% intergalactic magnetic fields which increases the effective particle
% pathlength compared to the light propagation distance to the cosmic
% ray sources and increases the energy loss. 
 For this purpose we extend the
 calculation of proton propagation in the local universe by Stanev {\it
 et al.} \cite{SEMPR} to cosmological distances.
 We also study the importance
 of the cosmological evolution of the sources of cosmic rays in different
 cosmological models. The aim of the present work is to study the level
 at which these ultra-high energy neutrinos are indeed guaranteed.

 Section \ref{sec2} discusses the neutrino production
 from propagation of ultra-high energy protons in the local Universe.
 In Section \ref{sec3} we obtain the neutrino spectra from homogeneously
 distributed cosmic ray sources accounting for the cosmological evolution
of these sources. Section \ref{sec4} explores variations in this flux 
under the influence of different assumptions concerning proton injection spectra, source
evolution and distribution, and background cosmology.
 Section \ref{sec5} gives a brief overview of the event rates that could 
 be expected in future
 large neutrino detectors. Discussion of the results and the conclusions
 from this research are given in Section \ref{sec6}.
 
\section{Neutrino fluxes from proton propagation in the local universe}
\label{sec2}

We begin with our method for calculating neutrino production from proton 
propagation in the the nearby universe (for a detailed discussion see 
\cite{SEMPR}). The calculation is carried out as a Monte Carlo simulation 
of individual particle histories in the presence of the cosmic background 
radiation, including energy loss processes such as photoproduction, 
$e^+e^-$ pair production, and adiabatic losses. The extension of this 
method to cosmological distances is discussed Sec. \ref{sec3}.

An important new ingredient of the calculation is to use the event 
generator SOPHIA ~\cite{SOPHIA} to simulate in detail the proton/neutron 
interactions with photons from the cosmic microwave background.
This event generator 
has several main differences from previously used codes:\\
 {\em (1)} the inclusion of direct pion production at the  
photoproduction threshold. In this $t$-channel process the photon is 
absorbed via a $\gamma \pi^+ \pi^+$ vertex, and so only charged pions are 
produced.  Although the cross section for this process is smaller than 
the dominant $\Delta^+$ resonance, it yields a  significant number of 
neutrinos, when folded with the steep  proton injection spectrum.\\ 
%
% {\em (1)} the representation of direct pion production at the
% photoproduction threshold - a process that only generates 
% charged pions. In terms of Feynman graphs, this process has 
% a strong vertex at the baryon branch and an electromagnetic vertex
% for the photon interaction. The presence of the electromagnetic
% vertex requires that the particle the nucleon couples to is charged.
% Although the cross section for this process is 
% smaller than the dominant $\Delta^+$ resonance, it yields a
% significant number of neutrinos, when folded with the steep
% proton injection spectrum.\\ 
%
{\em (2)} explicit consideration of 10 different resonance
production channels in the important energy region just above the particle
production threshold.\\
 {\em (3)} QCD motivated multipion production at large center-of-mass
energies.
%  - this improvement will not be very important for interactions
%  of the steep cosmic ray spectrum on microwave background, but can have
%  influence for hard photon spectra.

To evaluate the neutrino yields, protons are injected in narrow 
logarithmic bins and all products of their interactions are collected 
with the same energy binning. We use 10 bins per energy decade, and 10000 
protons per bin weighted by an $E^{-2}$ spectrum within each bin. The 
injection energy ranges from $10^{19}$ to $10^{23}$ eV. This gives the 
option to explore different injection power spectra and cutoff energies 
by rescaling the products of each energy bin. 

%%%%%%%%%%%%%%%%%%%%%%%%%%%%%%%%%%%%%%%%%%%%%%%%%%%
\begin{figure}[t] % fig 1a
\centerline{\epsfig{file=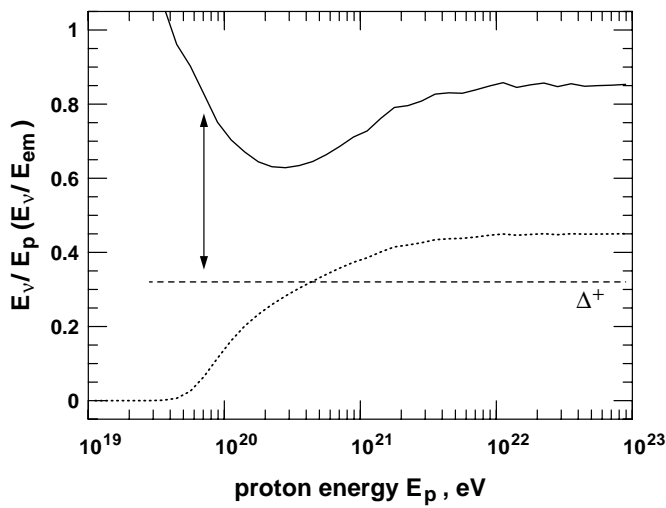,width=85mm}}
% \vspace{10pt}
\caption{Neutrino production efficiency, summed over flavors, as a function 
of proton injection energy. The solid curve shows
the ratio of the energy carried by neutrinos to that of
electromagnetic particles due to photoproduction
in fully developed cascades (200 Mpc), as simulated with SOPHIA
\protect\cite{SOPHIA}. The dashed curve is the same ratio 
but for the $\Delta^+$ resonance approximation which is 
frequently used in analytic calculations. The dotted curve shows the 
total neutrino energy relative to injected proton energy for SOPHIA.
\label{fig1a}
}
\end{figure}
%%%%%%%%%%%%%%%%%%%%%%%%%%%%%%%%%%%%%%%%%%%%%%%%%%%
The results from the Monte Carlo are illustrated in Fig.~\ref{fig1a}. 
Concerning the overall yield of neutrinos, the dominant feature is the 
turn on of the GZK process at $E_p \approx 5\times 10^{19}$ eV. The ratio 
of yield in neutrino energy to yield in radiative energy depends 
primarily on the ratio of charged to neutral pion production. If all pion 
production occurred through the $\Delta^+$ resonance this ratio would be
approximately $({{3} \over {4}} \times {{1} \over {3}}) / ({{2} \over {3}} + 
 {{1} \over {4}} \times {{1} \over {3}}) = {{1} \over {3}}$, 
 where for charged pions $\sim {{3} \over {4}}$ of 
 the energy goes to neutrinos. At high energies, isospin ``democracy"
 suggests that the ratio should tend to
 $( {{3} \over {4}} \times {{2} \over {3}}) / ({{1} \over {3}} + 
 {{1} \over {4}} \times {{2} \over {3}}) = 1.$
For low energy protons, direct production of charged pions plays an 
important role, again increasing the neutrino yield above that expected 
from the $\Delta^+$ resonance. 
%%The average neutrino energy is also plotted. 
%%In the $\Delta$ resonance model this number would be about 0.05, but one 
%%may see that SOPHIA yields a smaller value at threshold and that it decreases 
%%with increasing $E_p$.

The next step is to place the neutrino production model into an
astrophysical setting. For the proton source spectra we use a power law 
with an exponential high-energy cutoff
\begin{equation} 
 \frac{dN}{dE} \; \propto \; E^{-\alpha} \times \exp {(-E/\Ecut)}, 
\label{inj-spectrum}
\end{equation} 
where $\alpha$ = 2 unless otherwise stated 
and $\Ecut = 10^{21.5}$ eV. During propagation, adiabatic energy losses 
for the protons are calculated assuming $H_0$ = 75 km/s/Mpc. Similarly,
neutrino energies are redshifted by a factor of $(1+z)$, where $z$ is the 
redshift of the interaction site.

The energy degradation of ultra-high energy protons in propagation in the 
microwave background is very fast. The minimum mean free path for 
photoproduction interactions is 3.8 Mpc at a proton energy of 
$\sim$6$\times$10$^{20}$ eV. Protons with an energy of about 10$^{21}$~eV 
thus interact on the average twice or more during the first 10 Mpc of 
propagation and lose close to 50\% of their injection energy. A 
significant fraction of the energy loss (about 40\%) goes into neutrinos. 
The neutrino flux thus originates from the initial stages of proton 
propagation.

%%%%%%%%%%%%%%%%%%%%%%%%%%%%%%%%%%%%%%%%%%%%%%%%%%%
\begin{figure}[t] % fig 1
\centerline{\epsfig{file=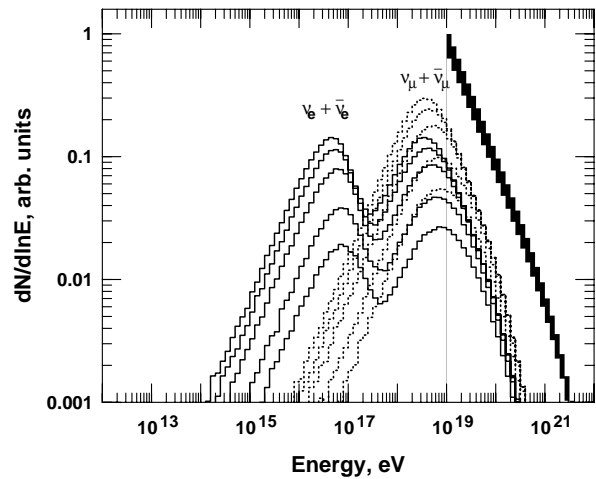,width=85mm}}
% \vspace{10pt}
\caption{Neutrino fluxes produced during the propagation 
of protons over 10, 20, 50, 100, and 200 Mpc
(from bottom up) in a 1 nG random magnetic field. The heavy histogram
shows the proton injection spectrum 
defined in Eq.~(\protect\ref{inj-spectrum}).
\label{fig1}
}
\end{figure}
%%%%%%%%%%%%%%%%%%%%%%%%%%%%%%%%%%%%%%%%%%%%%%%%%%%
 Fig.~\ref{fig1} shows the fluxes of electron and muon neutrinos
 after propagation over different distances up to a maximum of 200 Mpc.
About 60\% of the final neutrino fluxes are generated in the first 50 Mpc
and more than 80\% in the first 100 Mpc. The contribution from
 the second half of the maximum propagation distance is small because
 the proton spectrum is deprived of $> {\rm 10}^{20}$~eV particles
 and photoproduction interactions are rare. 
It follows, therefore, that from the point of view of neutrino production 
a source at 200 Mpc produces a fully evolved spectrum. Accordingly, for the 
cosmological scenarios that follow in Section \ref{sec3} we scale 
the neutrino yields to this result.

 There are two other noticeable
 features in the neutrino spectra shown in Fig.~\ref{fig1}. The muon neutrino
 spectra have a single peak at energies between 10$^{18}$ and 10$^{19}$ eV.
 Electron neutrinos, however, exhibit a more complicated double 
 peak structure. The first peak between 10$^{16}$ and 10$^{17}$ eV
 is populated by $\bar{\nu}_e$ from neutron decay. The neutron decay length
 equals the photoproduction interaction length at about 4$\times$10$^{20}$
 eV and neutrons of lower energy are more likely to decay than to interact.
 This leads to the formation of an additional peak in the electron neutrino
 spectrum. The second peak, in a position similar to that of muon neutrinos,
 is populated mostly by $\nu_e$ from $\mu^+$ decay with a small admixture
 of $\bar{\nu}_e$ generated predominantly in neutron photoproduction
 interactions. The ratio of $(\nu_\mu + \bar{\nu}_\mu)/(\nu_e + \bar{\nu}_e)$
 in the second peak is 2, as expected, although integrated over the
 whole spectrum the ratio is closer to 1.

%%%%%%%%%%%%%%%%%%%%%%%%%%%%%%%%%%%%%%%%%%%%%%%%%%%
\begin{figure}[t] % fig 2
\centerline{\epsfig{file=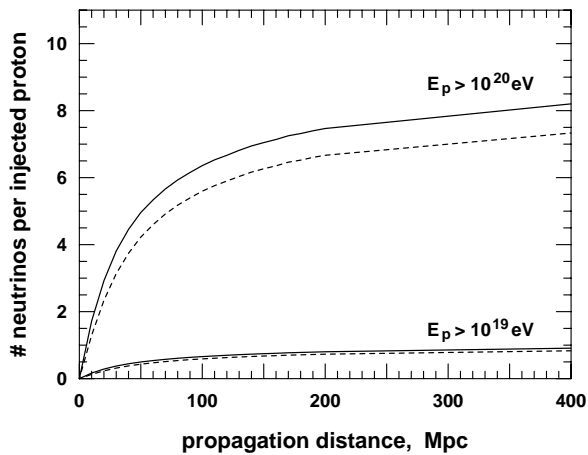,width=85mm}}
% \vspace{10pt}
\caption{Total number of neutrinos produced per injected proton of energy above
\protect$10^{19}$ eV (lower curves) and \protect$10^{20}$ eV (upper curves). 
The proton
energy was sampled from the spectrum (\protect\ref{inj-spectrum}). 
Solid lines show the sum of muon neutrinos and
antineutrinos, the dashed lines are for electron 
neutrinos and antineutrinos.
\label{fig2}
}
\end{figure}
%%%%%%%%%%%%%%%%%%%%%%%%%%%%%%%%%%%%%%%%%%%%%%%%%%%

One also notices the slight shift of the peak of the distribution to 
lower energy with the propagation distance. At longer propagation 
distance, lower energy protons suffer photoproduction interactions and 
generate lower energy neutrinos. There is also a small effect from the
adiabatic losses of all neutrinos, but this is hardly noticeable here 
because the maximum source redshift considered is only $z \approx 0.05$.

 Fig.~\ref{fig2}  shows the total number of
 neutrinos produced per source proton as a function of the source distance.
Because the protons lose most of their energy during propagation over
the first 50 Mpc,
 one would naively expect that the neutrino number does not change for
source distances above 100 Mpc.
The continuing rise of the neutrino to proton number
ratio is due to redshift effects. The minimum proton energy
for photoproduction interactions decreases as $(1+z)^{-1}$,
which leads to an increase of the number of interacting protons.
Even for relatively small redshifts involved ($z$ = 0.1 for
400 Mpc) this leads to an increase of the generated number 
of neutrinos.

 Fig.~\ref{fig2} also underlines the importance of the maximum
 energy of the proton injection spectrum. In this calculation
 we use an exponential cut--off with $\Ecut = 10^{21.5}$~eV.
 Assuming an $\Ecut$ of 10$^{20.5}$~eV would not drastically
 decrease the neutrino flux. 
 Cutting off the proton injection spectrum at lower energy would,
 however, require very nearby sources for
 the extremely high energy showers
 detected by the Fly's Eye and AGASA experiments.

%%%%%%%%%%%%%%%%%%%%%%%%%%%%%%%%%%%%%%%%%%%%%%%%%%%%%%%%%%%%%%%%%%%%%%%%

\section{Neutrinos from proton propagation over cosmological
 distances}
\label{sec3}

 In the following we will focus on the case of uniformly distributed 
 sources with identical proton injection spectra. Although a
 homogeneous source distribution is disfavored by the resulting 
 source energy requirements and arrival proton
spectra~\cite{BatSigl00,Olinto00}, it serves here as a simple generic
model whose results can be easily rescaled to account for 
local density enhancements or even nearby point sources.

The local neutrino flux of flavor $i$ generated from the propagation of
cosmic rays over cosmological distances can be written as an integral 
over redshift and the proton energy $E_p^s$ ($s$ denotes ``source")
\begin{equation}
 {\cal F}_i (E_{\nu_i}) = \frac{c}{4 \pi E_{\nu_i}} \int\!\!
 \int {\cal L}(z,E_p^s) Y(E_p^s,E_{\nu_i},z)
\frac{dE^s_p}{E^s_p}  dz .
\end{equation}
% 
% \begin{eqnarray}
% {\cal F}_i (E_{\nu_i})&=&\frac{dN_{\nu_i}}{d\ln E_{\nu_i} dt dA} 
% \nonumber\\
% & &\hspace*{-1cm} = 
% \int \frac{{\cal L}(z) }{(1+z) 4 \pi [R(0) f(r)]^2}
% \frac{dN_{\nu_i}}{dN_p d\ln E_{\nu_i}} d\ln E^s_p dV,
% \label{cos-flux}
% \end{eqnarray}
% 
% with the scale factor $R(z)$ and the volume element
% $dV=R^3(z)f^2(r)dr d\Omega$.
%
Here, the neutrino yield function is
\begin{equation}
Y(E_p^s,E_{\nu_i},z) = E_{\nu_i} \frac{dN_{\nu_i}}{dN_p dE_{\nu_i}}.
\end{equation} 
Also, the source function per unit redshift is
\begin{equation}
{\cal L}(z,E_p^s) = {\cal H}(z) \eta(z) {\cal L}_0(E_p^s) ,
\end{equation}
where ${\cal H}(z)$ parametrizes the cosmological source evolution, 
$\eta(z)$ describes the cosmological expansion, and
${\cal L}_0(E_p^s)$ is a properly normalized version of the source 
spectrum in Eq.~(\ref{inj-spectrum}).
The metric element $\eta (z)$ is defined as
\begin{eqnarray}
\eta (z) &=& \frac{dt}{dz} = {{1} \over {H_0 (1 + z)}}
\big[\Omega_M ( 1 + z)^3 + \Omega_\Lambda
\nonumber\\
& & + (1-\Omega_M-\Omega_\Lambda) (1+z)^2\big]^{-1/2}
\label{eta-factor}
\end{eqnarray}
which simplifies to $1/(H_0 (1 + z)^{5/2})$ for 
the Einstein -- de Sitter universe
($\Omega_M$ = 1, $\Omega_\Lambda$ = 0). 

The yield function $Y$ is evaluated utilizing the Monte Carlo result 
for a 200 Mpc source and the scaling relation 
\begin{equation}
Y(E_p^s, E_\nu,z)  = Y( (1+z) E_p^s, (1+z)^2 E_\nu,0).
\end{equation}
In scaling $E_\nu$ one factor of $(1+z)$ arises from redshifting 
the neutrino energy from its observed value to its production value. Both 
$E_\nu$ and $E_p^s$ scale by $(1+z)$ to maintain the same invariant 
reaction energies in the presence of a higher cosmic background 
temperature. Although it simplifies the numerical work considerably, 
utilizing the scaling relation introduces some approximations. For 
redshifts $\lapproxeq 0.05$ it overestimates the neutrino production as 
per Fig.~\ref{fig1}. However the contribution to the total fluxes coming
from $z<0.05$ is very small (see Fig.~\ref{fig3a} below). 
Another effect is that at high redshift the competition between 
neutron decay and neutron photoproduction is altered in favor of 
photoproduction, and so we make a modest overestimate of the 
$\bar{\nu}_e$ flux around $10^{16}$~eV. At high energies, the sum of 
$\bar{\nu}_e$ and $\nu_e$ fluxes remains unchanged, but the flavor 
distribution may be altered.
%% Third, for proton energies $(1+z) E_p^s$
%% in excess of $10^{22}$~eV, the maximum energy simulated, we perform an 
%%extrapolation of the neutrino yield based on approximate Feynman scaling 
%%of the hadronic interactions. 

The source proton luminosity ${\cal  L}_0$ is parametrized as 
\begin{equation}
{\cal  L}_0(E^s_p) = P_0 \left(\int_{E_{\rm min}}^{E_{\rm max}}
E^s_p \frac{dN_p}{dE_p^s} dE^s_p\right)^{-1 } E_p^s\frac{dN_p}{dE_p^s} ,
\end{equation}
with $dN_p/dE_p^s$ given by Eq.~(\ref{inj-spectrum}) and $P_0$ denoting
the injection power per unit volume. 

The injection
power of cosmic rays with energy above $E_{\rm min} = 10^{19}$ eV 
can be roughly estimated using the local cosmic ray energy density
\cite{Gaisser00c}.
The cosmic ray flux $dN/(dE d\Omega dA dt)$ at 10$^{19}$ eV
is about 2.5$\times$10$^{-28}$ cm$^{-2}$s$^{-1}$ster$^{-1}$GeV$^{-1}$.
Assuming that:\\
{\em 1)} all cosmic rays at that energy are extragalactic;\\
{\em 2)} 10$^{19}$ eV cosmic ray flux is as at injection; and\\
{\em 3)} the differential proton spectrum at injection is a power law
 with spectral index $\alpha$ = 2,\\
one obtains a cosmic ray energy density 
\begin{equation}
 \rho_e \; = \; \frac{4\pi}{c} \int E \frac{dN}{dE d\Omega dA dt} dE
\end{equation}
 of $1.1 \times 10^{54}$ erg/Mpc$^3$ per decade of energy. To calculate
 the injection power required to maintain this energy density one 
 needs to make an assumption about the lifetime $\tau_{CR}$ of these
cosmic rays. A conservative approach would be to use a lifetime
close to the Hubble time. Using $\tau_{CR} = 10^{10}$ yrs gives a power 
 of 1.1$\times$10$^{44}$ erg/Mpc$^{3}/$yr per decade of energy.
 Of course, the total power for $E >$10$^{19}$ eV depends on the maximum
 energy at acceleration.

The correct way of calculating the injection power for a model  of cosmic 
ray source distribution and injection (acceleration) spectra is to 
propagate the accelerated spectra from the sources to us and fit the 
locally observed spectrum. We do not perform this procedure because it 
involves assumptions on the cosmic ray source distribution and the 
structure and strength of the extragalactic magnetic fields which are 
beyond the scope of this paper. We use instead the cosmic ray injection 
power obtained in a similar, somewhat simplified way by 
Waxman~\cite{Waxman95}, who derived  $P_0 = 4.5 \pm 1.5 \times {\rm 
10}^{44}$ erg/Mpc$^3$/yr between 10$^{19}$ and 10$^{21}$ eV for power law 
cosmic ray injection spectra with $\alpha$ between 1.8 and 2.7.  We will 
use this value of $P_0$ for the energy spectrum of 
Eq.~(\ref{inj-spectrum}) integrated between 10$^{19}$ and 10$^{22}$ eV. 
The higher $E_{\rm max}$ approximately compensates for the factor of 
$\exp(-E/\Ecut)$ as compared to Waxman's result.

Finally, we have to specify the cosmological evolution of 
the cosmic ray sources, ${\cal H}(z)$. We use the parametrization
of \cite{Waxman95}, i.e.
\begin{equation}
  {\cal H}(z) =  \left\{
\begin{array}{lcl}
 (1 + z )^3  &  {\rm :}   &  z <  1.9\\
 (1 + 1.9)^3 &  {\rm :}   & 1.9 < z < 2.7\\
 (1 + 1.9)^3 \exp\{(2.7-z)/2.7\} & {\rm :} &
  z >  2.7
\end{array}\right. 
\label{evo} 
\end {equation}
where $n=3$ describes the source evolution up to moderate redshifts.
We also briefly consider a stronger evolution model with
$n=4$ up to $z$=1.9 and flat at higher redshifts. 

\begin{figure}[t] % fig 3
\centerline{\epsfig{file=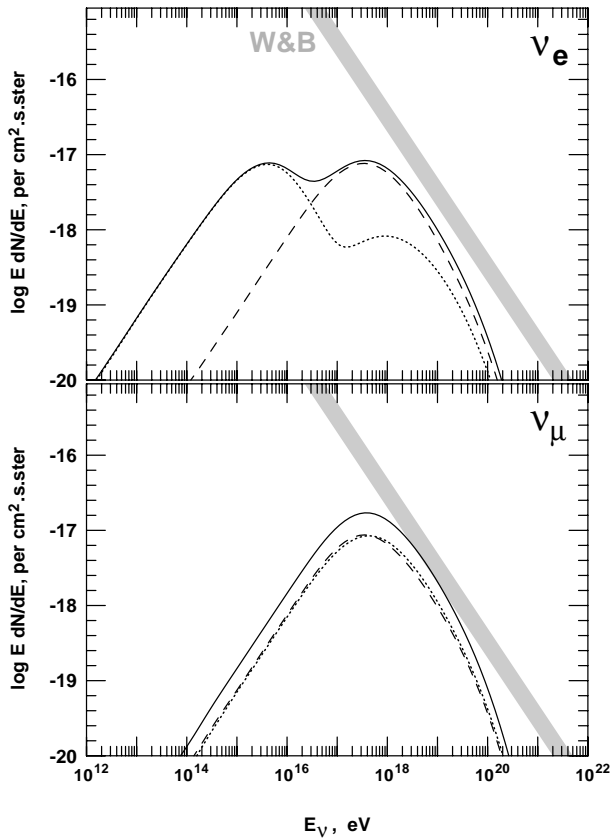,width=85mm}}
% \vspace{10pt}
\caption{Fluxes of electron neutrinos (dashed lines) and antineutrinos 
(dotted lines) generated in propagation of protons
are shown in the upper panel. The lower panel shows the
fluxes of muon neutrinos and antineutrinos. Solid lines show  the sum
of neutrinos and antineutrinos. The shaded band shows the Waxman \&
Bahcall~\protect\cite{WB1,WB2}
limit for neutrino production in cosmic ray sources with the
same injection power. The lower edge of the band is calculated without account
for the cosmological evolution and the upper one with the
evolution of Eq.~(\protect\ref{evo}).
\label{fig3}
}
\end{figure}

Fig.~\ref{fig3} shows electron and muon neutrino fluxes obtained with our 
nominal choice of astrophysical and cosmological parameters, and carrying 
out the integration to a redshift of $z_{\rm max} = 8$. Integrating to 
infinity increases the neutrino fluxes by only about 5\%.

Fig.~\ref{fig3} also shows the limits on neutrino production in cosmic 
ray sources derived by Waxman \& Bahcall~\cite{WB1,WB2} (WB). As those 
calculations were carried out for the same source evolution model, 
similar spectra and the same injection power $P_0$, they serve to compare 
the expectations for `source' versus `propagation' neutrinos associated 
with UHECR of astrophysical origin. Our propagation flux is slightly 
below the WB limit for the muon neutrino and antineutrino flux for 
energies between 10$^{18}$ and 10$^{19}$ eV. The differences lie in the 
assumed neutrino yield per proton. For their limit, WB assume a maximal 
thin source, i.e. an energy equal to that of the injected proton is 
deposited into neutrinos, whereas for our calculation only a fraction 
goes into neutrinos, as shown in Fig.~\ref{fig1a}. At higher energies one 
can see the effect of the factor $\exp (-E/\Ecut)$ in our source 
spectrum. At lower energies, the cosmic background radiation is devoid of 
high energy photons and so low energy protons do not produce neutrinos. 
In contrast, the cosmic ray sources are assumed to have abundant higher 
energy photons and so the limit on source neutrinos continues to scale as 
$E^{-2}$ to low energies.

% Fig.~\ref{fig3} shows electron and muon neutrino fluxes obtained with
% $n=3$ for a homogeneous source distribution.
% It compares the neutrino fluxes with the 
% Waxman \& Bahcall~\cite{WB1,WB2}
% limit on neutrino production in cosmic ray sources, which is based
% on the same injection power $P_0$. Although the
% conditions in the source are much different from intergalactic space, such
% a comparison shows the expectations for `source' versus `propagation'
% neutrinos associated with UHECR of astrophysical origin.
% The muon neutrino flux calculated above is similar to
% the Waxman \& Bahcall limit on the muon neutrino and antineutrino flux 
% for energies between 10$^{18}$ and 10$^{19}$ eV.
% The reason is obvious: the limit is derived for
% optically thin sources for the cosmic rays and the case that
% protons lose all their energy
%  in photoproduction interactions. 
%  Although in our calculation most of the protons 
%  with $E_p> {\rm 10}^{20}$ eV interact more 
%  than once in propagation over cosmological distances,
%  they only lose a fraction of their energy, as shown in Fig.~\ref{fig1a}.
%  Compared to the photon fields at active galactic nuclei (AGN) or
%  gamma--ray bursts (GRB) the microwave background photons are much
%  softer, which keeps the proton interaction threshold, and the
%  average neutrino energy, very high. 
% 
% The redshift integration in Fig.~\ref{fig3} 
% was carried out to a maximum redshift 
% of 8. An integration to infinity would increase
% ${\cal F}(E_\nu)$  only by about 5\%.
%%%%%%%%%%%%%%%%%%%%%%%%%%%%%%%%%%%%%%%%%%%%%%%%%%%
%%%%%%%%%%%%%%%%%%%%%%%%%%%%%%%%%%%%%%%%%%%%%%%%%%%

Fig.~\ref{fig3a} is designed to show how the neutrino flux is built up 
from contributions at different redshifts. It is evident that the high 
and low ends of the neutrino spectrum are sensitive to different epochs 
of the source evolution. First consider the protons that will contribute 
to neutrinos with energy $10^{19}$~eV. At $z=0$ these protons have an 
energy of a few times $10^{20}$ eV, above the threshold for photoproduction. 
This energy will increase with the source redshift. As a result, the 
source contribution $E_p dN/dE_p$ for these neutrinos effectively 
decreases as $(1+z)^{-1}$. To this we must add additional factors of 
 $\eta(z) {\cal H}(z) \sim (1+z)^{0.5}$ for the source evolution
 in a $\Omega_M=1$ cosmology, and a factor of $(1+z)$ explicit in 
the $(1+z) d/d(1+z)$ plot. Together, the function plotted naively scales 
as $(1+z)^{0.5}$. This scaling stops at $z=1.9$ where ${\cal H}(z)$ is 
assumed to flatten. For higher energy neutrinos $E_\nu= 10^{20}$~eV, the 
increasing proton energy runs into the exponential cutoff $\Ecut$ of our 
model injection spectrum causing a further decrease with $1+z$. The 
result of these considerations is that the highest energy neutrinos are 
produced primarily by relatively young sources, and are sensitive to 
assumptions about the recent universe.

For low energy neutrinos, say $10^{16}$~eV the story is a bit more 
 complicated. From kinematic arguments the prime production candidate
 for such neutrinos would be a 
proton of energy a few times $10^{17}$~eV, but such protons are below the 
photoproduction threshold. Protons with higher energy can, of course,
produce low energy neutrinos, but due to the small phase space the 
production is suppressed by a factor of $E_\nu/E_p$. Now, as the source 
redshift increases, $E_\nu$ at production also increases as $1+z$. At the 
same time, the minimum value for $E_p$ at production {\em decreases} due to 
the increasing cosmic microwave background
temperature. Thus, phase space considerations of the neutrino production 
process yield a net factor of $(1+z)^2$. With the lowering of $E_p$, the 
source spectrum factor yields an increase of $1+z$. Including $\eta(z) 
{\cal H}(z)$ and the explicit $1+z$ for the plot gives an overall 
dependence of $(1+z)^{4.5}$ at low energies. This behavior continues until\\ 
a) the source evolution model changes its $z$ dependence, or\\
b) the photoproduction threshold at $z$ has dropped so that there is no phase 
space suppression for that neutrino energy. At that point there is a 
transition to the high energy behavior outlined above. The net result of 
these considerations is that the low energy part of the spectrum is 
dominated by high redshift sources, and is sensitive to assumptions of  
a cosmological nature in our calculation.
%%%%%%%%%%%%%%%%%%%%%%%%%%%%%%%%%%%%%%%%%%%%%%%%%%%
\begin{figure}[t] % fig 3a
\centerline{\epsfig{file=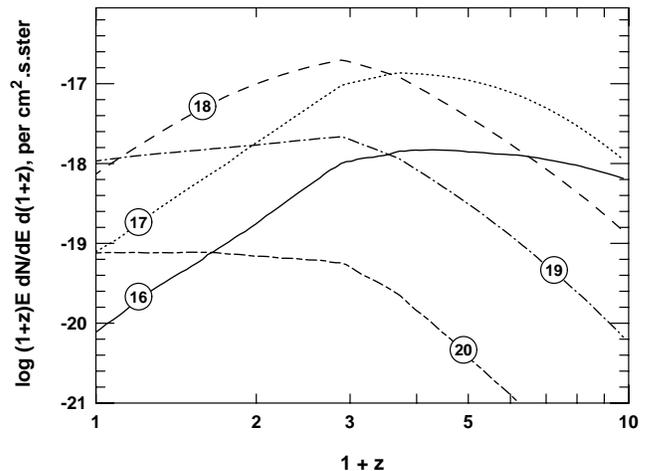,width=85mm}}
% \vspace{10pt}
\caption{The curves, labeled by $\log_{10}(E_\nu)$, show the 
contribution of different source distances to the neutrino flux 
as a function of redshift for our 
nominal $n=3$ source evolution model given in Eq.~(\protect\ref{evo}). 
\label{fig3a}
}
\end{figure}
%%%%%%%%%%%%%%%%%%%%%%%%%%%%%%%%%%%%%%%%%%%%%%%%%%%

Finally, we comment on the energy where the neutrino flux peaks in 
Fig.~\ref{fig3}. Given the turn on of photoproduction (Fig. \ref{fig1a}) 
and the kinematics of the $\Delta$ resonance, one might expect the peak 
to occur at around $10^{19}$ eV. Our Monte Carlo, however, 
yields more neutrinos with a softer spectrum than a $\Delta$ resonance 
model, so the peak from a low redshift source occurs at about $3 \times 
10^{18}$ eV, as seen in Fig. \ref{fig1}. Moreover, as discussed just 
above, the peak of the cosmological spectrum is shifted by two factors of 
$(1+z)$ from the redshift which dominates the source contributions. For 
our $\cal H$, $\eta$ and ${\cal L}_0$ this occurs at $(1+z) = 2.9$, and 
so the resultant neutrino spectrum peaks at around $3 \times 10^{17}$ eV 
as seen in Fig. \ref{fig3}.

%Assuming $\Delta$ resonance dominance, the peak of the neutrino spectrum 
%is expected to occur at a neutrino energy about a factor of 20 below the 
%photoproduction threshold at the epoch of strongest source evolution. For 
%our $\cal H$, $\eta$ and ${\cal L}_0$, and including the decrease in 
%$E_p$ with the photoproduction threshold, the source contributions peak 
%at $(1+z) = 2.9$, but are fairly flat for some range of $z$ above that 
%value. This yields an expected peak at around $10^{18}$~eV. The actual 
%peak seen in Fig.~\ref{fig3} is at about $3 \times 10^{17}$~eV. The shift 
%is due partly to the extended source contributions from high redshift and 
%partly to our detailed Monte Carlo, which produces more numerous but 
%softer neutrinos than a simple $\Delta$ resonance model.

%%%%%%%%%%%%%%%%%%%%%%%%%%%%%%%%%%%%%%%%%%%%%%%%%%%%%%%%%%%%%%%%%%%%%%%%

\section{Variations}
\label{sec4}

\begin{figure}[t] % fig 4
\centerline{\epsfig{file=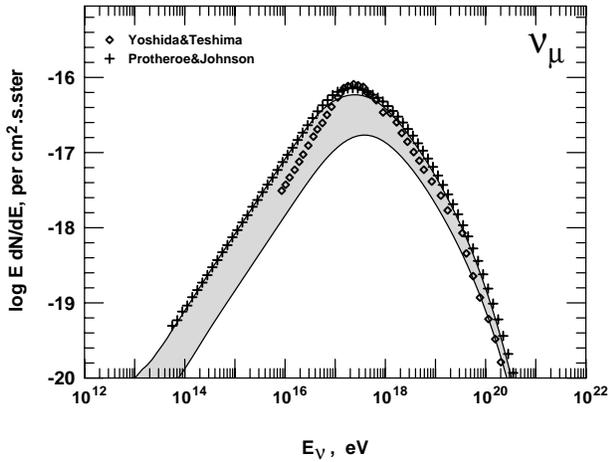,width=85mm}}
% \vspace{10pt}
\caption{ The lower boundary of the shaded area corresponds to
 the neutrino flux shown in Fig.~\protect\ref{fig3} 
with the source evolution 
\protect$n$ = 3 and the upper boundary is for \protect$n$ = 4
 up to \protect$z=1.9$ and constant afterwards. 
Open diamonds show neutrino fluxes calculated
by Yoshida and Teshima~\protect\cite{YT93} and the crosses
are due to Protheroe and Johnson~\protect\cite{PJ00}.
\label{fig4a}
}
\end{figure}

Many of the parameters associated with the calculation of the 
neutrino fluxes shown in Fig.~\ref{fig3} have rather large  
uncertainties. The power needed to maintain the flux of cosmic  rays 
above 10$^{19}$ eV varies by about 30\% for injection  differential 
spectral indices between 1.8 and 2.7~\cite{Waxman95}.  The cosmological 
evolution of the source luminosity evaluated  from star formation 
regions~\cite{Madau} could be somewhat stronger,  as also indicated by 
the attempts to derive the cosmological evolution of GRB~\cite{GRB_evo} 
and their fluences~\cite{Pugliese}. We show the influence on the 
generated neutrino fluxes in Fig.~\ref{fig4a} where we calculate the 
neutrino flux with the same injection power but a stronger cosmological 
evolution - $(1 + z)^4$ up to  $z$ = 1.9 and constant thereafter.  The 
stronger cosmological evolution increases the neutrino flux by  a factor 
of 3 and generates a small shift of the maximum flux  to lower energy. 
The integration was carried again to redshift of 8.

%%%%%%%%%%%%%%%%%%%%%%%%%%%%%%%%%%%%%%%%%%%%%%%%%%%
%%%%%%%%%%%%%%%%%%%%%%%%%%%%%%%%%%%%%%%%%%%%%%%%%%%

The two sets of points in Fig.~\ref{fig4a} represent calculations of 
Yoshida \& Teshima (diamonds, source evolution with $n$ = 4 and cutoff at 
$z=4$) and of Protheroe \& Johnson~\cite{PJ00}  (crosses, energy cutoff 
at $E= 10^{21.5}$ eV). 

%%Apart from
%% the different normalization, these two calculations show
%% peak neutrino energies different from the present calculation.
%% Yoshida \& Teshima predict a spectrum that peaks at lower energies.
%% We understand this difference in terms of the overestimate
%% of the proton photoproduction energy loss below 10$^{20}$~eV
%% compared to the current calculation \cite{SEMPR}. An additional
%% decrease of the
%% peak energy is related to the extension of the bright phase
%% of the cosmic ray sources to a redshift of 4.

 All three calculations show the peak of the neutrino spectrum 
 at approximately the same energy of 2-3$\times$10$^{17}$ eV.
 The spectrum of Yoshida \& Teshima is somewhat narrower then
 the one obtained in this work, while the agreement with 
Protheroe \& Johnson is very good. This latter work uses
 the cosmological evolution evolution model RLF2~\cite{Peacock}
 with the `fudge factor' of Rachen \& Biermann~\cite{RachBier}
 rather than a simple redshift dependence.

The neutrino flux calculated by Stecker {\em et al.} \cite{SDSS}
(not shown) seems to be based on an injection power not much different
from the normalization of Waxman.  The spectrum however peaks at higher
energy. An error might have been made in accounting for the neutrino
redshift (F.W.~Stecker, private communication).

%%%%%%%%%%%%%%%%%%%%%%%%%%%%%%%%%%%%%%%%%%%%%%%%%%%
\begin{figure}[t] % fig 4
\centerline{\epsfig{file=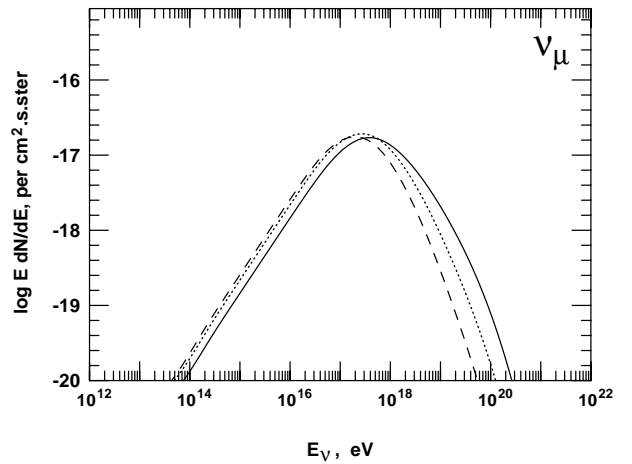,width=85mm}}
% \vspace{10pt}
\caption{Variation of neutrino flux resulting from different proton injection 
spectra. The dotted curve is for
an injection spectrum of \protect$E^{-2.5}$ and the dashed curve
 is for \protect$E^{-3}$.
\label{fig4b}
}
\end{figure}
%%%%%%%%%%%%%%%%%%%%%%%%%%%%%%%%%%%%%%%%%%%%%%%%%%%
In Fig.~\ref{fig4b} we show the neutrino fluxes obtained with the source 
evolution  of Eq.~(\ref{evo}) and differential injection spectral indices 
of 2.5 and 3, keeping again the injection power and $\Ecut$ constant.  
The steeper injection spectra generate smaller neutrino fluxes at high 
energy, because of the much smaller number of  protons above 10$^{20}$ eV 
that are mostly responsible for high energy neutrino production. By 
contrast, low energy neutrinos are 
predominately generated by protons injected at high redshift, where the 
threshold for photoproduction is decreased by a factor of $1+z$. 
Because the total injection power is kept constant,
a steeper spectrum results in an increase in the flux of 
lower energy protons and, hence, low energy neutrinos.

 All results shown above are calculated for a homogeneous source
 distribution. It has been suggested in the past,
 and recently in the context of a specific acceleration
 model in Ref.~\cite{BBO} that the observed cosmic ray
 spectrum can be best fit by a combination of a homogeneous 
 source distribution with an enhancement of local 
sources at distances less than 20 Mpc 
or with a single source at a similar distance.
% A realistic enhancement of the cosmic ray sources cannot exceed
% the local galactic overdensity of about a factor of 2.
 
%%%%%%%%%%%%%%%%%%%%%%%%%%%%%%%%%%%%%%%%%%%%%%%%%%%
\begin{figure}[t] % fig 5
\centerline{\epsfig{file=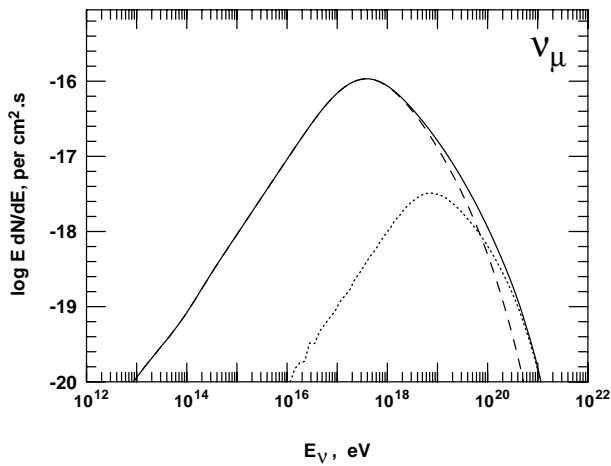,width=85mm}}
% \vspace{10pt}
\caption{ Muon neutrino and antineutrino fluxes generated if one
 half of the local
 cosmic rays are generated by nearby 
 sources at a distance of 20 Mpc. Note that fluxes are given for the full solid angle of
4\protect$\pi$. The dashed curve corresponds to
one half of the muon neutrino flux
of Fig.~\protect\ref{fig3}, integrated over the full solid angle. 
The dotted line shows the production
 by locally generated cosmic rays by either a single source or
 by a local enhancement of the cosmic ray sources. 
\label{fig5}
}
\end{figure}
%%%%%%%%%%%%%%%%%%%%%%%%%%%%%%%%%%%%%%%%%%%%%%%%%%%
Fig.~\ref{fig5} shows the neutrino fluxes generated under the
assumption that the injection power of homogeneously distributed
sources is $P_0/2$ and 50\% of the UHECR at $10^{19}$ eV are generated 
by a single source at a distance of 20 Mpc. This scenario will not
predict a proton arrival spectrum similar to that of Waxman
\cite{Waxman95} because the injection
spectrum is kept, up to the normalization, the same for the single 
source at 20 Mpc and the homogeneously distributed sources. 
It merely serves as a simple example of the changes to be expected in
such a case. It is also a good approximation for a local density
enhancement because
the difference between the neutrino flux magnitudes 
for a single source and local enhancement scenarios
is very small.

The observational difference is
substantial because, for a single source scenario, most of the
high-energy neutrinos
would come from the direction of that source. Most of the
neutrinos due to proton propagation from a local source would
be generated in the first interaction of protons 
of energy above 10$^{20}$ eV (see Fig.~\ref{fig2}). These high-energy protons 
do not scatter significantly in a random extragalactic field with an
average strength of less than 100 nG and the relativistic decay
kinematics ensures that the neutrinos are emitted in the direction of
the proton momentum.

Finally we discuss the importance of the cosmological model. All 
calculations shown above are performed with the assumption of a flat, mass 
dominated universe ($\Omega_M$ = 1). However, recent astrophysical 
observations agree better with models containing a cosmological constant 
$\Lambda$~\cite{NetaBetal}. From the behavior of $\eta$ 
(see Eq.~(\ref{eta-factor})) one may expect an increased contribution 
to the neutrino production from higher redshifts.

% Since these observations support cosmologies 
% in which the expansion accelerates, one expects an increased contribution 
% to the neutrino production from higher redshifts. {\bf huh?}

\begin{figure}[t] % fig 6
\centerline{\epsfig{file=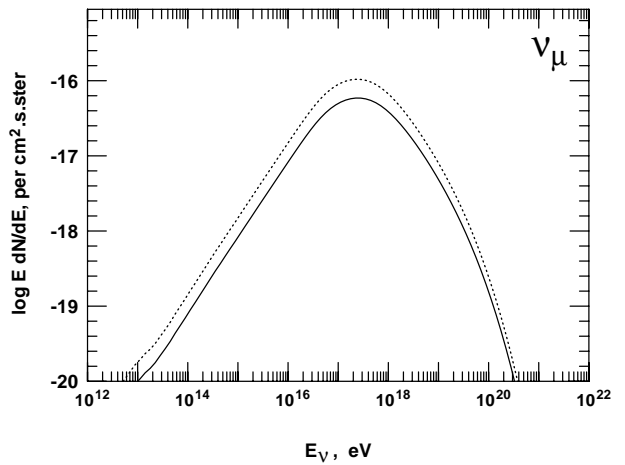,width=85mm}}
% \vspace{10pt}
\caption{Muon neutrino and antineutrino fluxes from
 homogeneous source distribution
 for the Einstein -- de Sitter Universe (solid line) and for
\protect$\Omega_\Lambda$ = 0.7 (dashed line), using a cosmological
evolution of \protect$(1 + z)^4$ up to redshift of 1.9 and flat
at higher redshifts. 
\label{fig6}
}
\end{figure}

Fig.~\ref{fig6} shows the difference in the expected neutrino fluxes for  
the Einstein -- de Sitter Universe (solid line) and a model with  
$\Omega_M$ = 0.3 and $\Omega_\Lambda$ = 0.7, as currently favored by 
measurements \cite{Carroll00a}, both with a source evolution  proportional to 
$(1 + z)^4$. We use the stronger source evolution and  carry 
the integration out to a redshift of 8  to emphasize the difference 
between the cosmological models. For a flat universe the ratio of fluxes 
is always smaller than $1/\sqrt{1-\Omega_\Lambda}$ (see 
Eq.~(\ref{eta-factor})).  The difference between the Einstein -- de 
Sitter Universe  and one with a non-vanishing cosmological constant does 
not, however, depend very strongly on the cosmological evolution  model 
of the sources. The ratio  between the two cosmologies is about 1.6 for 
$n$ = 3 evolution  and 1.7 for $n$ = 4 evolution. 

We caution that in the absence of a model which consistently accounts for 
the effects of $\Omega_\Lambda$ on the source evolution function ${\cal 
H}(z)$, this increase should be regarded as an upper limit. Specifically, 
we have varied $\eta$, but kept ${\cal H}$ constant, whereas it might be 
argued that keeping the product $\eta {\cal H}$ constant would be a 
better approximation to the effects of a non-zero $\Omega_\Lambda$ on 
the source 
evolution, in which case there would be no change in the neutrino flux.

%%%%%%%%%%%%%%%%%%%%%%%%%%%%%%%%%%%%%%%%%%%%%%%%%%%
%%%%%%%%%%%%%%%%%%%%%%%%%%%%%%%%%%%%%%%%%%%%%%%%%%%

%%%%%%%%%%%%%%%%%%%%%%%%%%%%%%%%%%%%%%%%%%%%%%%%%%%%%%%%%%%%%%%%%%%%%%%%%%

\section{Event rates from these neutrinos}
\label{sec5}

Detection of neutrinos produced during the propagation of UHECR protons 
is a challenge. The flux peaks above 10$^{17}$~eV where the neutrino 
nucleon cross-sections are of order 10$^{-31}$~cm$^2$. Such values of 
$\sigma_{\nu N}$ are large enough to make the Earth opaque, but still 
require 100-1000 km of water to ensure an interaction. As a result, there 
is no sensitivity to upward neutrinos and low efficiency for 
downward neutrinos. Under these circumstances the prime detector 
requirement is large mass, of order 100 km$^3$ (water), in order to 
guarantee a few events per year. With such large volumes, the 
typical event may be assumed to be a contained event where the visible 
energy is dominated by the high energy shower associated with the 
neutrino interaction vertex.  Special detector geometries may provide 
additional sensitivity to $\mu$ or $\tau$ leptons produced in charge 
current events. Although we phrase our discussion in terms of water/ice 
detectors, sufficient mass may also be achieved by 
monitoring large volumes of atmosphere.

All types of neutrino interactions generate showers. In charged current 
(CC) interactions of electron neutrinos and antineutrinos the neutrino 
energy is completely released in the form of shower particles. The 
hadronic shower carries a fraction $y$ of the initial neutrino energy 
$E_\nu$, while the electromagnetic shower carries the remaining $(1 - 
y)E_\nu$. Although the electromagnetic shower will be stretched out by 
the LPM effect\cite{AlvarezZ}, both showers are likely to be contained
 within the 
detector volume.  In charged current interactions of muon neutrino and 
antineutrinos  we only use $yE_\nu$ for the shower energy, as we do in the 
neutral  current (NC) interactions of all neutrino types. We use the 
GRV98~\cite{GRV98}  structure functions to calculate the neutrino cross 
sections  at ultra-high energy. The results are similar to those in 
Gl\"uck, Kretzer, and Reya\cite{GluckKR} and to those calculated by 
Gandhi~{\em et al.}~\cite{Gandhietal}  and Kwiecinski, Martin \& 
Stasto~\cite{KMS}.

%%%%%%%%%%%%%%%%%%%%%%%%%%%%%%%%%%%%%%%%%%%%%%%%%%%
\begin{figure}[t] % fig 7
\centerline{\epsfig{file=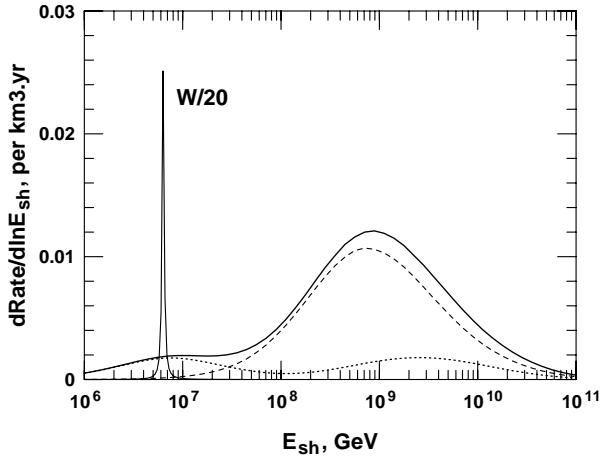,width=85mm}}
% \vspace{10pt}
\caption{ Differential rates for showers initiated by charge current
 interactions of electron neutrinos (dashes) and antineutrinos (dots).
 The solid line gives the sum of those and the dash--dot lines gives
the rate of the $W^-$ resonance events. Absorption by the Earth is 
included so that at high energy the solid angle for detection is $\Omega \approx 2 
\pi$.\\
\label{fig7}
}
\end{figure}
%%%%%%%%%%%%%%%%%%%%%%%%%%%%%%%%%%%%%%%%%%%%%%%%%%%
 Fig.~\ref{fig7} shows the differential rate for showers initiated by
 charged current interactions of electron neutrinos and antineutrinos for
a cosmological source evolution with $n$ = 3. The total shower
 rate is dominated by $\nu_e$ interactions around the peak energy
 of the neutrino flux. As the interaction cross section increases with 
energy, the signal is dominated by showers with energy above 10$^8$ 
GeV. At this energy there is no significant background, either from neutrinos 
or muons, produced by cosmic ray interactions in the atmosphere.
The spike at $E_\nu = 6.3 \times 10^6$ GeV is the rate for $W^-$ boson  
production by $\bar{\nu}_e$--electron interactions~\cite{Glashow_res},  
divided by 20. The differential rate in the vicinity of the  resonance is 
very high but the relevant energy range is small  and the total rate is  
0.03 per km$^3$ yr. Accounting for invisible  $W^-$ decay modes would 
reduce the rate for showers even further.

Table~\ref{tab-1} gives the shower rates per km$^3$ of water  per yr for 
showers generated by different types of  neutrino interactions and 
different flavors.  These rates represent $d\sigma_\nu/dy$  folded with 
the flux of neutrinos reaching the detector and do  not account for any  
experimental efficiency and detector biases. As such, they can only  
serve as an estimate of realistic event 
rates. The rates are low: the IceCube~\cite{ICECUBE} and
ANTARES~\cite{ANTARES} km$^3$-size 
experiments may not expect to detect these neutrinos. In addition violation 
of any of the following assumptions may further decrease expectations\\
(i) the detected ultra-high energy cosmic rays are not 
 produced by a single nearby source,\\
(ii) the powerful sources of ultra-high energy cosmic rays are
 homogeneously distributed in the Universe, and\\
(iii) our normalization of the
 power of the cosmic ray sources and the injection spectrum
 are correct.\\
In case that only a fraction of the UHECR comes
from homogeneously distributed sources, these rates come down
 to the same fraction of the values presented in Table~\ref{tab-1}. On 
the other hand, for
 the cosmological evolution model with $n$ = 4 the rates will
 go up by a factor of about 3, and assuming a cosmological
 constant with $\Omega_\Lambda$ = 0.7 could give another moderate
 increase by a factor of 1.6 -- 1.7.

%%%%%%%%%%%%%%%%%%%%%%%%%%%%%%%%%%%%%%%%%%%%%%%%%%%%%%%%%%%%%%
\begin{table}
\caption {Rates per km\protect$^3$ water per year of showers
above different energies  generated
 by different types of neutrino interactions for \protect$P_0 = 4.5
\times {\rm 10}^{44}$ erg/Mpc$^3$/yr
 and a cosmological evolution with \protect$n$ = 3 for homogeneously
 distributed cosmic ray sources (see text).  
\label{tab-1} }
\medskip
\begin{tabular} {l | c c c c c }
 log E\protect$_{sh}$ (GeV) \protect$>$ & 6 & 7 & 8 & 9 & 10 \\
\tableline
all \protect$\nu$, NC & 0.052 & 0.046 & 0.032 & 0.008 & 0.001 \\
\protect$\nu_e$, CC   & 0.054 & 0.051 & 0.046 & 0.024 & 0.004 \\
\protect$\nu_\mu + \nu_\tau$, CC & 0.092 & 0.080 & 0.057 & 0.014 & 0.002 \\
% \protect$\tau$ decay   & 0.30 & 0.30 & 0.26 & 0.12 & 0.02 \\
\tableline
total                 & 0.192 & 0.177 & 0.144 & 0.046 & 0.007 
\end{tabular}
\end{table}
%%%%%%%%%%%%%%%%%%%%%%%%%%%%%%%%%%%%%%%%%%

The rates shown in Table~\ref{tab-1} do not depend on the  assumption 
that the atmospheric neutrino anomaly could be explained by  
oscillations of $\nu_\mu$ into $\nu_\tau$~\cite{numunutau}.  On the very 
long  astrophysical pathlength 50\% of the muon neutrinos  would be 
converted to tau neutrinos. Charge current interactions  of $\nu_\tau$ 
would not, however, be different from the $\nu_\mu$ CC  interactions and 
so their sum does not depend on oscillation parameters. 

Another interesting process is the detection of $\tau$'s produced in
$\nu_\tau$ CC interactions in the material surrounding the detector.
Estimates show that the tau decay inside the detector will generate 
a signal which is comparable in shower energy and rate to that 
of $\nu_e$ CC interactions.
Whether or not $\tau$ decays can be observed as separate 
events depends on the $\tau$ energy, which affects both decay length and 
energy loss \cite{DuttaRSS}, and the detector volume. A proper study of event
rates for throughgoing or stopping $\tau$ and $\mu$ leptons depends on
the detector geometry, location and surrounding material and is beyond
the scope of this paper.

% 
% 
% This will be indeed true for the
%  highest energy $\tau$ mesons. The $\tau$ decay length at an energy 
%  of 10$^8$ GeV is 4.9 km, i.e. $\nu_\tau$ interaction and $\tau$
%  decay will not be observable in the same km$^3$ volume.
%  At lower energy, however, the two showers may look like
%  a single shower of higher energy, that would somewhat increase
%  the numbers shown in the table.
 
%%%%%%%%%%%%%%%%%%%%%%%%%%%%%%%%%%%%%%%%%%%%%%%%%%%%%%%%%

\section{Discussion and conclusions}
\label{sec6}

 The biggest uncertainty in the magnitude of the neutrino fluxes
 from proton propagation is related to the distribution of cosmic
 ray sources. All arguments about the normalization of the cosmic
 ray injection power and the cosmological evolution of the cosmic
 ray sources are based on the assumption that the detected UHECR
 are of astrophysical origin and are
 not accelerated at a single nearby source. If the latter were
 the case, and the nearby source were responsible for all particles
 above 10$^{19}$~eV, we should have to restrict the source distance
 to  less than 20 Mpc~\cite{SEMPR}. The local ultra-high
 energy cosmic ray density would then be much higher than the average
 cosmic ray density in the Universe. There will be, no doubts, other
 regions where ultra-high energy cosmic rays are accelerated
 and are over-abundant. The overall normalization of the cosmic ray
 power would then depend on the filling factor of such regions.
 If one estimates this filling factor from the volume of the walls
 of galactic concentration, such as the supergalactic plane, that
 describes the galactic distribution within a redshift less than
 0.05~\cite{lahav}, to nearby voids it will certainly not exceed
 10\%. In such a case the neutrino production due to proton propagation
 would be minimal as shown in Fig.~\ref{fig5} with the dotted line.

 A correct estimate of the power in UHECR depends on the strength
 of the extragalactic magnetic field in our neighborhood.
In case of an average strength of the turbulent field exceeding
1 nG, cosmic rays with an energy of about 10$^{19}$ eV and below
(with a gyroradius of less than 10 Mpc) will have diffusive
 propagation pattern, which will enhance their flux at Earth.
 On the other hand, regular extragalactic fields may guide
 these particles along the walls of matter concentration. All
 effects related to the propagation of UHECR should be a subject
 of further investigation before we could give a more reliable
 estimate of the UHECR power.

 Recently several authors have argued that powerful astrophysical
 systems and potential cosmic ray sources, such as GRB,
have a cosmological evolution stronger than 
$(1+z)^3$~\cite{Pugliese,Stecker2000}.
 Ref.~\cite{Pugliese} presents a combined analysis of the
 far infrared luminosity as a tracer of the star formation rate~\cite{Madau}
 as a function of redshift and the gamma ray burst fluence. The star
 formation rate is fit with an exponential rise of $\exp(2.9 z)$
 to a redshift of 1.7 and the GRB fluence distribution suggests a slow
 decrease at higher $z$. The use of such strong cosmological evolution
 and high activity at redshifts higher than 2 would place the
 estimate of the neutrino flux halfway between the two 
 cosmological evolution models shown in Fig.~\ref{fig6}.  

 Another important factor is the injection spectrum of UHECR  and 
the highest energy at acceleration. The qualitative  picture of its 
influence is demonstrated in Fig.~\ref{fig1a}. Generally protons have to 
be accelerated to  energies above 10$^{20}$ eV to generate significant 
neutrino fluxes from their propagation. This threshold is reduced by 
$(1+z)$ for contributions from high redshift. In view of the 
observations of cosmic  rays of energy significantly higher than 
10$^{20}$ eV this is very  likely if the sources of these particles are 
astrophysical objects.

Assuming homogeneously distributed astrophysical sources we 
obtain a neutrino flux similar to the Waxman \& Bahcall limit
at neutrino energies above 10$^{18}$~eV. This means that, with the
assumptions and restrictions discussed above, one may expect
similar fluxes of UHE neutrinos produced in astrophysical 
sources and in UHECR propagation.

Since the signature of these ultra-high energy neutrinos are showers, 
different types of air shower detectors should also be  able to observe 
the highest end of the neutrino spectrum.  The effective volume of the 
Auger observatory for UHE neutrino  interactions was estimated to 30 
km$^3$ of water equivalent~\cite{CroZas}.  If this effective volume was 
achieved for a shower energy of  10$^{19}$~eV, the Auger observatory 
would see about 0.3 events  per year from the estimates shown in 
Table~\ref{tab-1}. It is interesting to note that at an energy of 10$^{18}$ 
eV the $\tau$  decay length ($l_\tau = 49 E/(10^{18} {\rm eV})$ km) is of 
the order of the  dimensions of the Auger observatory. `Double bang' 
events,  caused by $\tau$ neutrinos as suggested by Learned \&  
Pakvasa~\cite{LeaPak}, could be detected if the sensitivity of the array 
was significant in this energy range.

As the effective volumes required are extremely large, the proposed 
satellite air shower experiments EUSO and OWL~\cite{EUSO,OWL} might be 
well suited for the observation of ultra-high energy neutrino fluxes; 
however, with the current advertised threshold of $5 \times 10^{19}$~eV 
most of the potential event rate would go undetected. Because of their 
large field of view, these detectors should in principle be able 
to observe `double bang' events of energy above 10$^{19}$ eV -- if the 
threshold energy were lowered to that level.

Showers generated by ultra-high energy neutrinos could also be observed 
by their radio emission \cite{ZasHS,FrichterRM}. Prototype experiments are 
in operation \cite{RICE} and suggestions have been made for full scale 
experiments \cite{SeckelF,Seckel} that would have an energy threshold of 
10$^{18}$~eV and a full effective volume of 10$^2$-10$^4$ 
km$^3$. Such detectors could take advantage of the higher shower rate 
that corresponds to the maximum of the neutrino flux as shown in 
Fig.~\ref{fig7}.

  The potential detection of ultra-high energy
 neutrinos is a crucial experimental result that will
 help us distinguish between  an astrophysical (acceleration)
 and cosmological (top--down) origin of UHECR.
 In top--down scenarios the neutrino fluxes are primary,
 roughly equal to the gamma ray fluxes and at least
 an order of magnitude above the ultra-high energy nucleon
 fluxes. In all astrophysical scenarios the neutrinos,
 due to cosmic ray interactions at their sources or in propagation,
 are secondary and their flux is a fraction of the cosmic ray
 flux.
 
Measuring neutrinos from CR propagation 
can also help to distinguish between protons
and heavy nuclei, such as iron, as highest energy cosmic rays.
The energy loss of heavy
nuclei during propagation over cosmological distances is governed
by photo-disintegration. The absorption of photons leads mainly to 
giant dipole resonance excitation of the nuclei and, with a high probability, 
subsequently to the emission of a single nucleon in the de-excitation 
process~\cite{RachenPhD}. 
Hence the neutrino spectrum is expected to be dominated by
relatively low-energy neutrinos ($E_\nu \lapproxeq 10^7$ GeV)
from the beta decay of neutrons and unstable nuclei.

Finally it should be mentioned that 
the magnitude of the flux and the arrival direction of UHE neutrinos are
a good indication of the cosmic ray source distribution in
astrophysical scenarios.

%%%%%%%%%%%%%%%%%%%%%%%%%%%%%%%%%%%%%%%%%%%%%%%%%%%%%%%%%%%%%%%%%%%%%%%%%%%%%%%%

\section*{Acknowledgments}
 The authors acknowledge helpful discussions with T.K.~Gaisser,
 R.J.~Protheroe and F.W.~Stecker.
 The research of TS is supported in part by NASA Grant NAG5-7009.
 RE is supported in part by the US Department of Energy contract
 DE-FG02 91ER 40626. These
 calculations are performed on DEC Alpha and Beowulf clusters funded
 by NSF grant PHY--9601834.

%%%%\ifonecol\raggedright\fi
%
\bibliographystyle{prsty}
%\bibliography{uhecr,physics}

\end{document}